\documentclass[preprint, aps, floatfix, showkeys,showpacs]{revtex4}
\usepackage[english]{babel}
\usepackage[latin1]{inputenc}
\usepackage{graphicx}
\usepackage{babel}
\makeatother
\begin{document}
\title{Synchronization in the presence of memory}
\author{Rafael~Morgado$^{1,2}$, Micha\l{}~Cie\'sla$^{1}$,
Lech~Longa$^{1,2}$,
 and Fernando~A.~Oliveira$^{2}$\linebreak}
\affiliation{$^1$Marian Smoluchowski Institute of Physics,
Jagellonian University, Department of Statistical Physics and Mark
Kac Complex Systems Research Center,\\ Reymonta 4, Krak\'{o}w, Poland}
\affiliation{$^2$Institute of Physics and International Center of
Condensed Matter Physics, University of Bras\'ilia, Campus
Universit\'ario Darcy Ribeiro, CP 04513 - CEP 70919-970 Bras\'ilia -
DF, Brazil}
\date{\today}
\begin{abstract}
 We study the effect of memory on  synchronization of identical
chaotic systems driven by common external noises. Our examples show
that while in general synchronization transition becomes more
difficult to meet when memory range increases,  for intermediate
ranges the synchronization tendency of systems can be  enhanced.
Generally the synchronization transition is found to depend on the
memory range and the ratio of noise strength to memory amplitude,
which indicates on a possibility of optimizing synchronization by
memory. We also point out on a close link between dynamics with
memory and noise, and recently discovered synchronizing properties
of networks with delayed interactions.
\end{abstract}
\maketitle
    Since a generic study of Fahy and Hamman  \cite{lit-fahy}
synchronization of dynamical systems has become an  active field of
research. Examples of synchronous behavior are found in physical,
biological, chemical and social systems \cite{lit-acebron}, from a
road traffic anticipation \cite{lit-treiber}, population growth
\cite{lit-grenfell} and secure communications
\cite{lit-communication} through biophysics \cite{lit-neuron},
chemistry \cite{lit-kiss} and laser optics \cite{lit-deshazer} to
computer science \cite{lit-computers}. At least four types of
synchronization scenarios have been identified \cite{lit-synchro} of
which the synchronization between identical systems coupled by a
common noise has received much attention due to its relative
simplicity and importance \cite{lit-identical}.

    In recent years studies of synchronization were extended further
to obey an important case of interactions that are delayed in time
\cite{lit-schuster,lit-bouchner,lit-delayed}.  One of the most
striking observations here was enhancement of synchrony for networks
of many nonlinear interacting units by time-delayed transmission of
the signals \cite{lit-delayed,lit-dhamala,lit-atay,lit-masoller}.
More specifically, in a neural network model \cite{lit-dhamala} an
enhanced synchronization of neurons by delays has been observed.
Similar effect was found for a network of coupled logistic maps
\cite{lit-atay}. Even a network of logistic maps with random delay
times was able to sustain synchronization \cite{lit-masoller}.

 A purpose of this work is to show that the constructive influence
of delays on synchronization can be extended further to obey a broad
class of nonlinear systems with noise and memory, where the latter
is understood as the auto-feedback with delay having distribution in
time. A close link between the synchronizing networks under delay
and synchronization by a common noise of (effectively decoupled)
nonlinear systems with memory is also demonstrated. Detailed
numerical calculations are carried out for generalized logistic maps
and chaotic Fahy-Hamman systems described by non-Markovian Langevin
equations. In both cases the choice of  the models has been
motivated by their well documented synchronizing properties in the
limit of vanishing memory. Importantly, our analysis demonstrates
that the presence of memory and noise not only can sustain
synchronization existing in equivalent, memoryless systems, but also
can  enhance it. The models chosen, though governed by different
dynamics, collectively display this possibility, which suggests that
the effect can be quite common for nonlinear interacting systems.

Memory and randomness naturally link with time evolution of
interacting dynamical systems
\cite{Zwanzig,lit-morgado1,lit-lee-ergodicity}. Indeed, the
evolution of a system coupled with 'external' degrees of freedom
(\emph{e.g.} open systems, nonlinear networks) can, at least in
principle, be reduced to a dynamics of an effective single system,
but with memory and noise.  The effective system is usually more
amenable to numerical, analytical and formal considerations. Perhaps
the simplest and exact example of such reduction is that of a
nonlinear system coupled bi-linearly to harmonic oscillators. The
elimination of the oscillator degrees of freedom results in a
generalized Langevin equation for the dynamics of the system, where
memory and noise terms are fully specified by the properties of the
oscillators and by their coupling with the system \cite{Zwanzig}.
Similarly, the dynamics of a network with delay times can be
approximated by, or in some cases reduced to an effective dynamics
of single nodes with memory and noise. Consider, for example, a
network of coupled logistic maps with discrete time
\cite{lit-atay,lit-masoller}. The state $x_i(t+1)$ of the node $"i"$
at time $t+1$ depends on the state $x_i(t)$ of that node at time $t$
and on the states $x_j(t-\tau_{ij})$ of the nodes \{"$j\ne i$"\}
that couple to $"i"$ at earlier times \{t-$\tau_{ij}$\}, where
$\{\tau_{ij}\}$ are the delay times (see \emph{e.g} Eq. (1) in
\cite{lit-atay,lit-masoller}). By iterating the equations for
$x_j(t-\tau_{ij})$ ($j\ne i$) back in time and re-substituting them
to $x_i(t+1)$ we arrive at the effective, single-node equations for
$x_i(t+1)$ expressed in terms of $x_i(t)$ and the nonlinear
auto-feedback (memory)  $\Gamma$. For the network of logistic maps
$\Gamma$ is a polynomial in $x_i(t-n),\, 1 \le n \le t$ with
coefficients depending on network's connectivity and the initial
values $x_j(0)$ for the nodes. Any randomness in the original
network like random delay times, random elements in network's
connectivity, or averaging over $x_j(0)$ goes into
additive/multiplicative noise terms in the effective equations of
motion.

One of the simplest, but important, class of $\Gamma$s is a linear
auto-feedback with $\Gamma \sim \sum_{k=1}^{N}\Gamma_{k}x_i(t-k)$,
which for example, can represent a 'mean-dynamics' of the above
mentioned networks of logistic maps. For $\Gamma$ given by a
polynomial in $x_i(t-n)$ a recipe  for getting $\{\Gamma_{k}\}$ of
the 'mean-dynamics' would be \emph{e.g.} a replacement of $x_i(t-n)$
by $x_i(t-n)= <x> + [x_i(t-n)-<x>]$ $ = <x> + \delta x_i(t-n)$ and
neglect  of terms that are nonlinear in $\delta x_i$; $\left\langle
\cdot \right\rangle $ denotes the average over the trajectory and
over initial conditions.

 The discussion as given clearly shows that memory and noise are
intrinsic to dynamical evolution of a system. It is then important
to know in what way they  affect synchronization. We report on the
noise-induced synchronization, which is generic for this case. The
synchronizing system is characterized by master trajectories that
divide the whole phase space onto basins of attraction such that all
trajectories initiated in the same basin and subjected to the same
sequence of the noise evolve to the same master trajectory
\cite{lit-fahy,lit-computers}.

As our first model we consider an ensemble of  chaotic logistic maps
coupled by common, additive noises. Since the coupling is realized
only through the noise terms it is sufficient to explore just two
such systems, which we define as:
\begin{equation}
    x_{n+1}^i = 4x_{n}^i(1-x_{n}^i)
+I\sum_{k=1}^{N}\Gamma_{k}x_{n-k}^i +\xi_{n} +\epsilon^i_n \mbox{
\hspace{1mm} mod } 1,
    \label{eq-nonsymetric}
\end{equation}
where $i = 1,2$ and $\xi_{n}$ is the non-symmetric,
$\delta$-correlated noise taken uniformly from the interval
$[a,a+b]$. To avoid roundoff-induced synchronization we also add an
extremely small independent uniform noise $\epsilon^i_{n}$ (from
interval $[-10^{-12},10^{-12}]$) to each particle at every iterated
step \cite{lit-longa}. We extend our analysis to a symmetric version
of (\ref{eq-nonsymetric}) by defining a new variable
\begin{equation}
z_{n}^i=x_{n}^i-\left\langle x\right\rangle ,\label{eq-symetric}
\end{equation}
where $\left\langle x\right\rangle\equiv \left\langle
x_n^i\right\rangle$.  For $I=0$ the noise-induced synchronization in
the $[a,b]$-plane has recently been studied in detail by Rim
\emph{et al.} \cite{lit-13rim}. We restrict ourselves to linear
auto-feedback with $\Gamma=I\sum_{k=1}^{N}\Gamma_{k}x_{n-k} $.
 The set of coefficients $\{I \Gamma_{k}\}$  is the 'memory profile' and  $N$ is
proportional to the memory range. Two models for the memory profile
are considered in detail: the constant memory profile with
\begin{equation}
\Gamma_{n} \equiv \Gamma^{c}_{n}=\left\{ \begin{array}{cc}
1 & \hspace{0.5cm} \mbox{for}\,\, \hspace{0.5cm} n\leq N,\\
0 & \hspace{0.5cm} \mbox{for}\,\, \hspace{0.5cm}
n>N,\end{array}\right. \label{eq-constant-memory}
\end{equation}
and the exponentially decaying  memory profile
\begin{equation}
\Gamma_{n} \equiv \Gamma^{e}_{n}=\left\{ \begin{array}{cc}
\exp(-\lambda n) & \hspace{0.5cm} \mbox{for}\,\, \hspace{0.5cm} n\leq N,\\
0 & \hspace{0.5cm} \mbox{for}\,\, \hspace{0.5cm}
n>N.\end{array}\right.\label{eq-exponential-memory}
\end{equation}
For  $N\gg 1/\lambda$ the inversion of  $\lambda$ is the memory
range.

The stability of the synchronized states is determined by the sign
of the (maximal) transversal Lyapunov exponent $\Lambda$ for the
dynamics of difference $\delta
x_{n}=x_{n}^1-x_{n}^2=z_{n}^1-z_{n}^2$. In the numerical
calculations we choose at random  the initial states $\{
x^1_{N},x^1_{N-1},..., x^1_{0} \}$ from the allowed interval and
nearby states $\{ x^2_{N},x^2_{N-1},..., x^2_{0} \}$. Then we
iterate the equations of motion to construct statistics of the
expansion and contraction rates: $\lambda_i=\ln\left(\frac{|\delta
x_{N+i+1}|}{|\delta x_{N+i}|}\right)$ of $\delta x_{n}$. The
procedure, repeated for many randomly chosen initial states, allows
us to calculate the average of $\lambda_i$, which approximates
$\Lambda$ \cite{lit-bouchner,lit-dhamala,lit-13rim}.

In case of  non-zero memory we generally find that for large enough
absolute noise intensity, $|I|$, the synchronization is destroyed
for all $N\ge 1$. Results are shown in Fig. \ref{fig-Treshold} where
$|I|$, above which the synchronization region disappears, is
sketched. For $|I|$ exceeding the threshold value the systems
de-synchronize for all $a$ and $b$. Choosing $\lambda=\frac{2}{N}$
we find that the results compare well for both memory profiles. In
this case $\Gamma^{e}_{N}$ at $n=N$ is about an order of magnitude
smaller than at $n=0$ ($\Gamma^{e}_{N}=e^{-2}$).
\begin{figure}[htbp]
    \begin{picture}(240,160)
    \put(-20, 160){
      \includegraphics{fig1.ps}
    }
    \end{picture}\vspace{-10mm}
\caption{Threshold lines for symmetric map (\ref{eq-symetric}): (a)
$\Gamma^{e}_{n}$, $I<0$, (b) $\Gamma^{e}_{n}$, $I>0$, (c)
$\Gamma^{c}_{n}$, $I<0$, and (d) $\Gamma^{c}_{n}$,  $I>0$. For
$\Gamma^{e}_{n}$ cases we take $\lambda=\frac{2}{N}$.
\label{fig-Treshold}}
\end{figure}
The effect is illustrated further  in Fig. \ref{fig-shrinking},
where the evolution of the synchronizing boundaries,
$\Lambda(a,b)=0$, are shown with increasing (positive) intensity for
the symmetric map (\ref{eq-symetric}) and  for different memory
profiles. The case without memory \cite{lit-13rim} is also shown for
comparison. Please note that the synchronization area shrinks with
increasing intensity and range of the memory. This behavior is
observed for  $N\ge 1$, for positive and negative intensities, and
for all maps studied.
\begin{figure}[htbp]
    \begin{picture}(240,160)
    \put(-20, 160){
      \includegraphics{fig2.ps}
    }
    \end{picture}\vspace{-10mm}
\caption{Synchronization areas for symmetric map. Region (a)
corresponds to system without memory \cite{lit-13rim}. Left plot is
done for $\Gamma^{c}_{n}$ with (b) $N=5$, $I=0.1$ and (c) $N=5$,
$I=0.3$. Right plot corresponds to $\Gamma^{e}_{n}$ with
$\lambda=\frac{2}{N}$ and with (d) $N=5$, $I=0.1$ and (e) $N=5$,
$I=0.5$. \label{fig-shrinking}}
\end{figure}
Interestingly, the maxima in Fig. \ref{fig-Treshold} prove  that
\emph{memory can also act  on synchronization in a  constructive way
by enhancing it}. We observe  the enhancement of synchrony by memory
for both memory profiles, which indicates that the phenomenon can be
quite general. Indeed, as demonstrated  below  systems with a more
complex dynamics, governed by the integrodifferential, generalized
Langevin equation, show similar behavior. We discuss the possible
implications of these results toward the end.

Memory profile is closely linked to a time-time autocorrelation
function
\begin{equation}
C_{k}=\frac{\left\langle \left(x_{n}-\left\langle x\right\rangle
\right)\left(x_{n-k}- \left\langle x\right\rangle
\right)\right\rangle }{\left\langle \left(x_{n}-\left\langle
x\right\rangle \right)^{2}\right\rangle }. \label{eq-correlation}
\end{equation}
We monitored the behavior of this function inside- and outside of the
synchronizing area. Exemplary results are shown in Figs.
\ref{fig-correlation-functions-constant} and \ref{fig-Decay-time}.
The decay time $\tau$ in Fig. \ref{fig-Decay-time} was calculated by
assuming that  $C_{k}$ is a linear combination  of trigonometric
functions multiplied  by an exponential decay of the form
$\left|C_{k}\right|=\exp\left(-\frac{k}{\tau}\right)$. We observe in
Figs. \ref{fig-shrinking} and \ref{fig-Decay-time},  that close to
the border of the synchronizing area $\tau$ is enhanced by the
presence of memory, with the maximum positioned outside this area.
That is, the chaotic systems with  larger $\tau$ 'remember for
longer' about their initial conditions, which makes synchronization
more difficult and explains intuitively the shrinkage of the areas
in Fig. \ref{fig-shrinking}.
\begin{figure}[htbp]
    \begin{picture}(240,160)
    \put(-20, 160){
      \includegraphics{fig3.ps}
    }
    \end{picture}
\caption{Correlation functions for the symmetric map with constant
memory ($C^c_{\,\,\,k}$) and without memory ($C^0_{\,\,\,k}$). Here
$N=5$, $I=0.1$, $b=0.1$, $a=5.3$ ( $a=5.55$ for insets).
\label{fig-correlation-functions-constant}}
\end{figure}
\begin{figure}[htbp]
    \begin{picture}(240,160)
    \put(-20, 160){
      \includegraphics{fig4.ps}
    }
    \end{picture}
\caption{Decay time of  $C_k$ as a function of noise parameter $a$.
Cases studied are: (a) constant memory and  (b) exponential memory
for $I=0.1$ and $\lambda= 2/N$. The case (c) corresponds to I=0. The
remaining parameters are $N=5$ and $b=0.1$.\label{fig-Decay-time}}
\end{figure}

The second model with which we explore influence of memory on
noise-induced synchronization is the generalization of the
Fahy-Hamman system \cite{lit-fahy}. We analyze trajectories of two
identical particles in a two dimensional potential well given by:
\begin{equation}
    V(x_1,x_2) = \frac{\sin 2\pi x_1}{2 \pi x_1} +
    \frac{\sin 2\pi x_2}{2 \pi x_2} +
        \frac{(x_1^2 + x_2^2)^2}{16 \pi^2},
\end{equation}
with different initial conditions. The motion of the $i$-th particle
($i=1,2$) is governed by the generalized Langevin equation:
\begin{equation}
\label{eq-eqmotion}
    m \ddot{x}^i_\alpha(t) = -\frac{\partial V(x^i_1,x^i_2)}{\partial
    x^i_\alpha} -
        m I \int^t \Gamma(t-t') \; \dot{x}^i_\alpha(t') \; dt'
        + \xi(t),
\end{equation}
where $m$ is the mass, $I$  is again the memory intensity or, in
this case,  the friction constant. The noise $\xi$,  common for both
particles, is a $\Gamma$-correlated stochastic force with zero mean,
where correlations are obeying fluctuation-dissipation theorem:
\begin{equation}
   \langle \langle\, \xi (t) \xi (t') \,\rangle \rangle =
   2 m I k_B T \; \Gamma(t-t'),
\label{eq-fd}
\end{equation}
with $T$ being the absolute temperature and $k_B$  the Boltzmann
constant.
 In what follows we restrict ourselves to the exponentially
correlated noise by choosing, as in Eq.
(\ref{eq-exponential-memory}),  $\Gamma(t-t') = e^{-\lambda
(t-t')}$. Double angular brackets denote averaging over a noise
realization.  The equations of motion (\ref{eq-eqmotion}) are
integrated numerically using stochastic version of the  Euler
algorithm. Discretization of the equations (\ref{eq-eqmotion})
entails  re-scaling the noise strength by a factor $1/\sqrt{\Delta
t}$, where  $\Delta t$ is the  time step. Finally, the exponentially
correlated noise is generated  from uniformly distributed random
numbers through Ornstein-Uhlenbeck process. As previously,
simulations are carried out to determine the maximal Lyapunov
exponent, $\Lambda$, as function of  memory range ($1/\lambda$) and
$I$. We  used a natural system of units: energy $\epsilon_u = V(0,0)
- V_{min} \approx 2.41$, time $t_u=\sqrt{3}$ measuring curvature of
the potential at origin and length $l_u = 1$ giving the period of
the oscillating part of the potential. In the absence of memory the
system was originally studied by Fahy and Hamman \cite{lit-fahy}
using regular Andersen thermostat. The results unambiguously showed
that trajectories were  exponentially convergent to a common
trajectory after a transient period. The same phenomenon has been
reported for a Langevin dynamics without memory of a one-dimensional
Lennard-Jones chain \cite{lit-4ciesla} and of other systems
\cite{lit-computers}.

Calculations of $\Lambda$ in the presence of memory as function of
 $1/\lambda$ are shown in Fig.~\ref{fig-ll}.
Please note that the dependence of $\Lambda$ on $1/\lambda$ is much
more complex now than that previously observed for the maps.   At
least three regimes can be identified.
 In the first regime, corresponding to  a short memory range, we observe
de-synchronization of the system by memory. However, after reaching
maximum, $\Lambda$ drops down and  for intermediate memory range
synchronization is considerably enhanced. The strongest
synchronization conditions are met  for $1/\lambda \approx 0.14$,
where $\Lambda$ approaches minimum. Interestingly, the positions of
the extremes are practically independent on $I$.
\begin{figure}[htbp]
    \begin{picture}(240,170)
    \put(-10, 170){
      \includegraphics{fig5.ps}
    }
    \end{picture}
 \caption{Maximal Lyapunov exponent $\Lambda$ against
memory strength $1/\lambda$ for three different friction constants
$I$ and the reduced temperature $k_B T = 1$.\label{fig-ll} }
\end{figure}

Summarizing, the results obtained for the generalized  logistic maps
and for the dynamical system evolving according to the generalized
Langevin equation uncover  a possibility of having the
\emph{constructive} influence  of memory on the noise-induced
synchronization. The results are quite counterintuitive for, in the
first place, we would expect that memory by introducing extra
dimensions \cite{lit-farmer} should act in just the opposite way,
{\emph{i.e.}} making synchronization more difficult
\cite{lit-atay,lit-masoller}. Though the proposed models are
relatively simple they represent quite different dynamics, which
suggests that the uncovered enhancement of synchrony by memory can
be a general phenomenon, occurring for a wide class of nonlinear
dynamical systems with memory and noise. They rise a possibility of
seeking for a right memory profile to create optimal conditions for
synchronization to occur \emph{ i.e. optimizing synchronization by
memory}. Additionally, our findings apart from giving yet another
example of the nontrivial interplay between memory and noise, show a
close relation to the synchronizing properties of coupled networks
with delays that have recently been discovered.

This work was supported by Brazilian CNPq Grant No~201388/03-0,
Grant N202 169 31/3455 of Polish Ministry of Science and by the EC
Marie Curie Host Grant for Transfer of Knowledge "\emph{COCOS}". We
also gratefully acknowledge the computing grant G27-8 towards the
use of the  ICM computers (University of Warsaw) and would like to
thank dr P. G\'{o}ra for useful discussions.

\end{document}